# The physical and mechanical properties of hafnium orthosilicate: experiments and first-principles calculations

Zhidong Ding[1], Mackenzie Ridley[2], Jeroen Deijkers[2], Naiming Liu[2], Md Shafkat Bin Hoque[1], John Gaskins[1], Mona Zebarjadi[2,3], Patrick Hopkins[1,2,4], Haydn Wadley[1,2], Elizabeth Opila[1,2], Keivan Esfarjani[1,2,4,*]

[1]*Department of Mechanical and Aerospace Engineering, University of Virginia, Charlottesville, Virginia 22904, USA*

[2]*Department of Material Science and Engineering, University of Virginia, Charlottesville, Virginia 22904, USA*

[3]*Department of Electrical and Computer Engineering, University of Virginia, Charlottesville, Virginia 22904, USA*

[4]*Department of Physics, University of Virginia, Charlottesville, Virginia 22904, USA*

\* Corresponding author.

E-mail address: k1@virginia.edu (K. E.)

# Abstract

Hafnium orthosilicate (HfSiO$_4$: hafnon) has been proposed as an environmental barrier coating (EBC) material to protect silicon coated, silicon-based ceramic materials at high temperatures and as a candidate dielectric material in microelectronic devices. It can naturally form at the interface between silicon dioxide (SiO$_2$) and hafnia (HfO$_2$). When used in these applications, its coefficient of thermal expansion (CTE) should match that of silicon and SiC composites to reduce the stored elastic strain energy, and thus risk of failure of these systems. The physical, mechanical, thermodynamic and thermal transport properties of hafnon have been investigated using a combination of both density functional theory (DFT) calculations and experimental assessments. The average linear coefficient of thermal expansion (CTE) calculated using the quasi-harmonic approximation increase from $3.06 \times 10^{-6}$ K$^{-1}$ to $6.36 \times 10^{-6}$ K$^{-1}$, as the temperature increases from 300 to 1500 K, in agreement with both X-ray diffraction lattice parameter and dilatometry measurements. The predicted thermal conductivity from Boltzmann transport theory was approximately 18 W/m.K at 300K. Both hot disk and laser flash measurements gave a thermal conductivity of 13.3 W/m.K. This slightly lower value is indicative of residual disorder in the experimental samples that was absent in the theoretical analysis. First-principles calculations and nanoindentation techniques were used to assess the ambient temperature elastic constants and bulk modulus respectively. The elastic properties obtained by both approaches agreed to within 5% validating the computational approach and its future use for study of the thermomechanical properties of other oxides or silicates.

## 1. Introduction

Compounds of Group-IV elements have many applications in a wide variety of fields. For instance, orthosilicates such as $ZrSiO_4$, $HfSiO_4$, $ThSiO_4$ and $USiO_4$, are effective radiation-resistant materials that are hosts for plutonium during the dismantling of nuclear weapons [1,2]. Zirconium- and hafnium-based borides, carbides and nitrides possess extremely high melting points, high hardness, but modest oxidation resistance, and may replace silicon-based ceramics for temperatures above 1700 °C [3,4]. Other applications of the group IV oxides arise in thermal barrier and the environment barrier coatings (TBCs and EBCs). For example, yttrium stabilized zirconia is used as a thermal barrier coating material that is applied to gas turbine components because it is stable, has low thermal conductivity and is conveniently prepared using air-plasma-spraying [7–10] or vapor deposition technology[11] . The group IV silicates have also shown potential as TBC or EBC materials. Ueno *et al* examined corrosion behaviors of $ZrSiO_4$ (zircon) and $HfSiO_4$ (hafnon) EBC materials in a water vapor environment at 1500 °C on silicon nitride specimens [13]. Although hafnon exhibited higher silica volatility, the underlying substrate experienced noticeably less oxidation and crack propagation. This was thought to be a consequence of hafnon's coefficient of thermal expansion (CTE) match with the silicon nitride susbstrate. Hisamatsu *et al* [14] created an EBC configuration that placed a hafnon layer between an yttria-stabilized hafnia topcoat and a ceramic substrate, and thus protected the substrate interface from cracks arising from stress.

Many of the late Group IV oxides have shown better gate dielectric performances than silicon dioxide in field-effect transistors, due to their large dielectric constants, stability in direct contact with silicon, and low leakage current in the ultrathin film regime [5,6]. Hafnium silicates have also been investigated as possible high dielectric candidates to replace $SiO_2$ gate dielectrics due to their phase stability, large bandgap, and compatibility with microelectronic fabrication processing environments [15–18]. Wilk *et al* obtained a dielectric constant of 11 from their hafnium silicate samples, which corresponds to an equivalent oxide thickness of 17.8 Å [15]. The high thermal stability of hafnium silicate is also of interest since it provides resistance to the thermal aging problems encountered with many dielectric gates [16].

The physical properties of hafnon have been investigated by both experiments and theoretical calculations. Its Raman-active fundamental frequencies [19–21] and its linear coefficient of thermal expansion (CTE) [22,23] were measured. First-principles [24] and molecular dynamics (MD) simulations [25] have provided consistent phonon properties of hafnon with the experiments. However, the Gruneisen parameters of hafnon calculated by a variety of simulation approaches [25–28] have not converged, and no experimental data is currently available for comparison. The thermal transport properties and mechanical properties of hafnon have also not been fully investigated by either theory or experiment.

Reliable data for the physical and mechanical properties of hafnium silicate are also needed to assess the growing list of applications. Here we conduct a comprehensive assessment of the thermal and mechanical properties of hafnon using both first-principles calculations and a variety of experiments. Starting with its lattice structure determined by Speer *et al* [29], the phonon dispersion, Gruneisen parameters, CTE, thermal conductivity, and elastic constants and moduli are calculated. Measurements of the CTE of hafnon *via* X-ray diffraction (XRD) and dilatometry, thermal conductivity through hot disk and the laser flash techniques, and the elastic modulus and hardness by the nanoindentation measurements are reported, and compared with our predictions and the results of other studies.

## 2. Methods

### 2.1 Physical and mechanical property calculations

The Quantum Espresso (QE) package [30,31] was used for structural relaxation and force constant calculations with the finite displacement approach. We applied a revised Perdew-Burke-Ernzerhof (PBEsol) [32] exchange-correlation functional with the GBRV-ultrasoft pseudopotential [33] that treats Hf ($5s$, $5p$, $5d$, $5f$, $6s$ and $6p$), Si ($3s$ and $3p$) and O ($2s$ and $2p$) as valence states. For structural optimization and self-consistent calculations, a cutoff energy of plane wave expansion is set to 100 Ry, and the reciprocal space is sampled by a 4×4×4 Monkhorst-Pack k-mesh [34]. A force convergence criterion is set as $10^{-3}$ eV/Å for tests on the exchange-correlation functional, the pseudopotential, the cutoff energy and the k-mesh. Harmonic force constants (FCs) are fitted by the PHONOPY package [35] and the ALAMODE package [36]. For phonon dispersion and second-order force constants calculations, 2×2×2 supercells with atomic displacements of 0.02 Å and a 2×2×2 Monkhorst-Pack k-mesh are generated by PHONOPY to ensure the convergence. We also apply force calculations on 88 configurations of 2×2×2 supercells with random displacements to fit cubic FCs using ALAMODE. As hafnon is a polar material, to include the non-analytical correction term [37] that leads to the splitting between longitudinal optical (LO) and transverse optical (TO) modes, the Born charge correction was applied to the phonons, Gruneisen parameters and thermal conductivity calculations.

Gruneisen parameters were calculated according to the definition.

$$\gamma_{qj} = -\frac{\partial \log \omega_{qj}}{\partial \log V} = -\frac{(e_{qj}^*)^T \delta D(q) e_{qj}}{6\omega_{qj}^2} \quad (1)$$

Here $V$ is the unit cell volume, $\gamma_{qj}$ is the Gruneisen parameter for each $q$ point for the $j$th phonon band, $\omega_{qj}$ and $e_{qj}$ are the corresponding phonon frequency and eigenvector, $\delta D(q)$ is the change in the dynamical matrix due to a 1% lattice parameter (i.e. 3% volume) increase. It can be either obtained from the difference in dynamical matrices calculated with the lattice parameters $a$ and $1.01\ a$; or from a Taylor expansion of the harmonic force constants in first power of atomic displacements: $\Phi_{ij}(u) = \Phi_{ij}(0) + \Sigma_k \Psi_{ijk} u_k + O(u^2)$ where the atomic displacement $u$ corresponds to a uniform 1% lattice constant expansion. In this work, we used the latter approach and later validated it by comparing to the results from the former one. Cubic FCs $\Psi_{ijk}$ are used to approximate $\delta D(q)$. To obtain reliable cubic FCs, it is important to determine the cutoff radius for each pair of atoms, because considering all the pairs would significantly increase computational cost. Besides, long-distance pairs may introduce noise to the fitting and affect the fitting quality. Therefore, when using ALAMODE, for each triplet of atoms in the fitting of cubic FCs, only Hf-O-Hf, Hf-O-Si, O-Hf-O and O-Si-O bonds were kept, and all non-bond triplets were excluded. To calculate the Gruneisen parameters, a q-mesh of 30×30×30 was used to assure convergence of the sums over the first Brillouin zone. Finally, with the fitted cubic FCs and the harmonic phonon frequencies and normal modes obtained from PHONOPY, the Gruneisen parameters of hafnon were calculated using the thirdorder.py script and the ShengBTE code [38].

The volumetric thermal expansion of hafnon was calculated using the quasi harmonic approximation (QHA) approach that is implemented in PHONOPY [39]. Eight conventional cells were created with varied volumes around the equilibrium structure and used to calculate the lattice dynamical properties of each structure. For each temperature, a map of Helmholtz free energies at different volumes was obtained. By fitting the free-energy versus volume curves at each temperature between 300 and 1500K, the corresponding equilibrium volume could be found, and thus the volume expansion and volumetric CTE were deduced. For a better comparison to experimental measurements from XRD, we also calculated the linear CTE of hafnon, $\alpha$, according to the relation between the Gruneisen parameter and linear CTE for anisotropic materials:

$$\alpha_i = (\partial \epsilon_i / \partial T)_\sigma = \Sigma_{j=1}^{6} C_\epsilon S_{ij}^T \gamma_j \tag{2}$$

In this formula, $\sigma$ and $\epsilon$ are respectively the stress and strain tensors, $T$ the temperature, $C_\epsilon$ the constant-strain volumetric heat capacity, $S_{ij}^T$ the isothermal elastic compliance tensor, and $\gamma_j$ the average Gruneisen parameter, and $(i, j)$ which vary from 1 to 6 are cartesian components in Voigt format. For hafnon that is a tetragonal system, once we define two average Gruneisen parameters: $\gamma_1 = \gamma_2 \neq \gamma_3$, we obtain the linear CTE $\alpha_1$ and $\alpha_3$ along the a- and c- axes, respectively as:

$$\alpha_1 = C_\epsilon[(S_{11}^T + S_{12}^T)\gamma_1 + S_{13}^T\gamma_3] \tag{3}$$

$$\alpha_3 = C_\epsilon[2S_{13}^T\gamma_1 + S_{33}^T\gamma_3] \tag{4}$$

The thermal conductivity of hafnon was calculated by solving the Boltzmann transport equation within the relaxation-time approximation (RTA) using ALAMODE. The phonon lifetimes were obtained through the phonon linewidth corresponding only to three-phonon scattering processes. This leads to a decrease of the thermal conductivity proportional to 1/T at high temperatures.

## 2.2 Mechanical property predictions

For elastic moduli calculations, Lagrangian strains were applied to the tetragonal unit-cell which contains 24 atoms (see Fig. 2 and captions) to generate deformed structures using the ElaStic code [40]. Total energy of several strained structures, are fitted with a second-order polynomial in powers of strain in order to extract the elastic constants. A cross-validation (CV) method was then adopted to evaluate the fitting quality. A fourth- or sixth-order polynomial was chosen for fitting and the variation of the elastic constants was kept below 0.1 GPa. The Voigt bulk ($B_V$) and shear ($G_V$) moduli were calculated from the elastic stiffness constants $c_{ij}$ and compliances $S_{ij}$ assuming that the strain was uniformly applied everywhere [41]:

$$B_V = \frac{1}{9}[(c_{11} + c_{22} + c_{33}) + 2(c_{12} + c_{13} + c_{23})] \tag{5}$$

$$G_V = \frac{1}{15}[(c_{11} + c_{22} + c_{33}) - (c_{12} + c_{13} + c_{23}) + 3(c_{44} + c_{55} + c_{66})] \tag{6}$$

The Reuss bulk, $B_R$ and shear moduli, $G_R$ were obtained using, [41].

$$B_R = [(S_{11} + S_{22} + S_{33}) + 2(S_{12} + S_{13} + S_{23})]^{-1} \tag{7}$$

$$G_R = 15[4(S_{11} + S_{22} + S_{33}) - (S_{12} + S_{13} + S_{23}) + 3(S_{44} + S_{55} + S_{66})]^{-1} \tag{8}$$

The Hill bulk, $B_H$ and shear, $G_H$ moduli were determined as the average of Voigt and Reuss moduli [41].

$$B_H = \frac{1}{2}(B_V + B_R) \qquad (9)$$

$$G_H = \frac{1}{2}(G_V + G_R) \qquad (10)$$

Once the bulk modulus and the shear modulus were calculated, Young's modulus $E$ and Poisson's ratio $\nu$ were obtained from the linear elastic solid approximation [42]:

$$E = \frac{9BG}{3B + G} \qquad (11)$$

$$\nu = \frac{3B - 2G}{2(3B + G)} \qquad (12)$$

## 2.3 Experimental lattice parameter and CTE measurements

Hafnon samples were made from 20-60 µm Praxair grade #02-P6644SG granules composed of sub-micron $SiO_2$ and $HfO_2$ particles as shown in Fig. 1a. Equimolar mixtures of the powders were loaded into a graphite die of 20 mm diameter for consolidation via spark plasma sintering (Thermal Technology LLC SPS Model 25-10). The SPS process was conducted using a maximum pressure of 65 MPa and maximum temperature of 1650 °C reached using a heating rate of 150 °C/minute. Samples were held at the maximum temperature and pressure for 40 minutes prior to cool down. Final bulk material underwent a 1300 °C heat treatment for 24 hours to restore oxygen stoichiometry, remove any residual carbon and remove possible stresses from processing. The resulting microstructure is shown in Fig. 1b. Phase analysis of samples was conducted by XRD (X-Ray Diffraction, Panalytical X'Pert Diffractometer) and SEM (Scanning Electron Microscopy, FEI Quanta 650F). Density measurements were made using Archimedes' Principle. An average relative density of 97.6% was achieved, assuming a theoretical density for hafnon of 6,970 kg/m$^3$. Samples were polished to one µm with diamond with a final polish using 0.05 µm colloidal silica in preparation for SEM and XRD measurements. Room temperature XRD results plotted in Fig. 1c as the log of intensity showed the presence of a small amount of remnant hafnia within the processed material. Residual hafnia peaks are those boxed in red, while all others correspond to $HfSiO_4$. The hafnia content is estimated to be less than 5 volume percent based on XRD, and analysis of plan view SEM micrographs where the lighter phase in Fig. 1b represents $HfO_2$ inclusions.

To determine the coefficient of thermal expansion via dilatometry, each sample was cut into a 15x3x3mm rectangular bar for dilatometer testing, using a Netzsch Dil 402c dilatometer. It was then heated in flowing argon at a rate of 3°C per minute with 60 recordings of length change per minute. The sample was held for 15 minutes at temperature before the cooling phase was started. The CTE was also determined from changes in lattice parameter during heating of the samples in an Anton Parr HTK 1200N non-ambient X-ray diffractometer. In this approach, the sample was radiation heated from room temperature to 1200 °C at 60 °C/min. The temperature was held constant every 100 °C to capture an X-ray scan of 15 – 60 degree 2-theta range. Rietveld refinement and lattice determinations were computed for each XRD plot using the HighScore Plus Software [Malvern Panalytical, Malvern, UK]. From these data, direction-dependent thermal expansion coefficients were determined for the given temperature range.

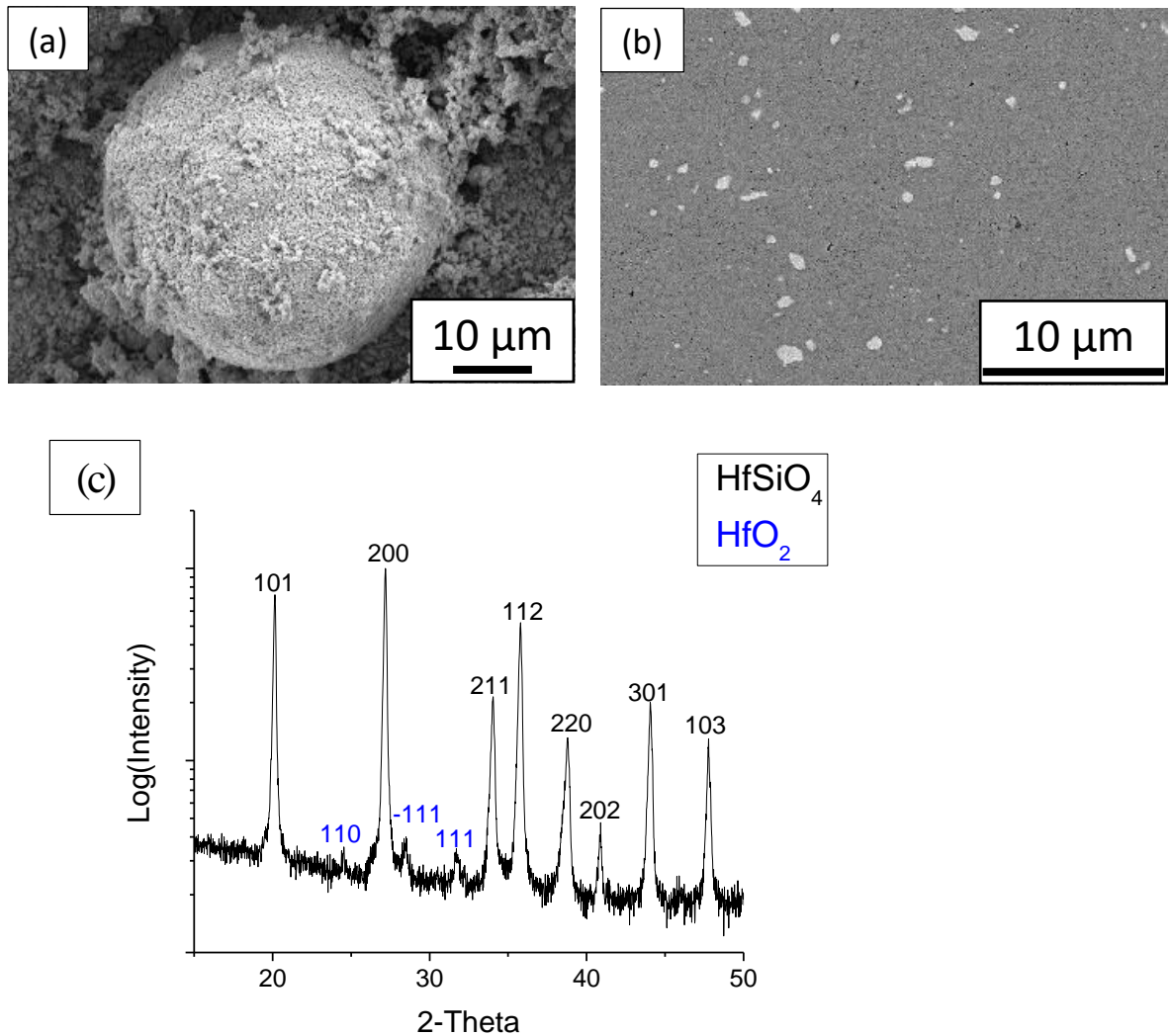

Figure 1. (a) Example of the $SiO_2$ and $HfO_2$ blended powder granule in the as received condition. (b) The hafnon sample microstructure after spark plasma sintering showing the presence of a small volume fraction of unreacted hafnia inclusions (the lighter contrast regions estimated to be 2.4% volume fraction), and pores (the darker contrast regions). (c) Room temperature XRD where remanents of $HfO_2$ peaks were also identified.

## 2.4 Thermal conductivity and elastic property measurement techniques

A hot disk approach using a (TPS 3500, Hot Disk AB) was used to measure the thermal conductivity of a pair of hafnon samples with thickness of 6 mm and diameter 20 mm from 30 to

300 °C in a box furnace CARBOLITE GERO 30 – 3000 ºC. Hot disk uses the transient plane source (TPS) method [43,44] to determine thermal conductivity. Two different sensors were used, one with a kapton layer and one with a mica layer over the nickel sensor. Prior to measurements, both sensors were used to measure a stainless steel standard (13.8 ± 0.5 W/m-K) to literature values of approximately 14 ± 0.7 W/m-K [45]. A laser flash technique was also used to determine the thermal diffusivity from which the thermal conductivity could be deduced. For the thermal diffusivity measurement conducted using a laser flash apparatus (467 HyperFlash, NETZSCH), a 10 mm×10 mm×3.5 mm size sample was cut. The density of the sample was measured by the Archimedes' method. The thermal conductivity $\kappa$ was then obtained from $\kappa = \rho c_p \alpha$ where $\rho = 6.97 \, g/cm^3$ is the theoretical density and $c_p$ is the specific heat, which is also temperature dependent.

The elastic modulus and hardness of the hafnon were measured with a nanoindenter (MTS XP) using standard continuous stiffness measurement (CSM) procedures from literature [46]. Prior to measurements on hafnon, the modulus and hardness of a silica standard were measured and compared with literature values (values in brackets). These tests gave a modulus of 72 ± 2 GPa (73.8 ± 0.3 GPa [47]) and hardness of 9.7 ± 0.4 GPa (8.85 ± 0.05 GPa [48]).

## 3. Results and discussion

### 3.1 The lattice structure

Our XRD measurements revealed that the hafnon samples have a body-centered tetragonal (zircon-like) structure with a space group of $I4_1/amd$ (No. 141). The conventional (tetragonal) unit cell consists of four formula units as shown in Figure 2. Highly symmetric hafnium, silicon and oxygen atoms coordinates in units of lattice parameters a and c are (0, 3/4, 1/8), (0, 1/4, 3/8) and (0, u, v), respectively, where u and v are internal parameters given in Table 1. As shown in Fig. 2, hafnium atoms are at the center of decahedra (purple), and silicon atoms are at the center of tetrahedra (yellow), and oxygen atoms at the corners are in red. Each tetrahedron shares two edges with adjacent decahedra. The lattice constants are the same along the a- and b- axes, which are slightly larger than that along the c-axis. Our optimized lattice constants are summarized in Table 1 and compared with experiments [29] and other calculations [24–28]. The difference in the lattice parameters predicted by the PBEsol method were within <1 % of measurements, while the other simulation methods gave larger relative errors.

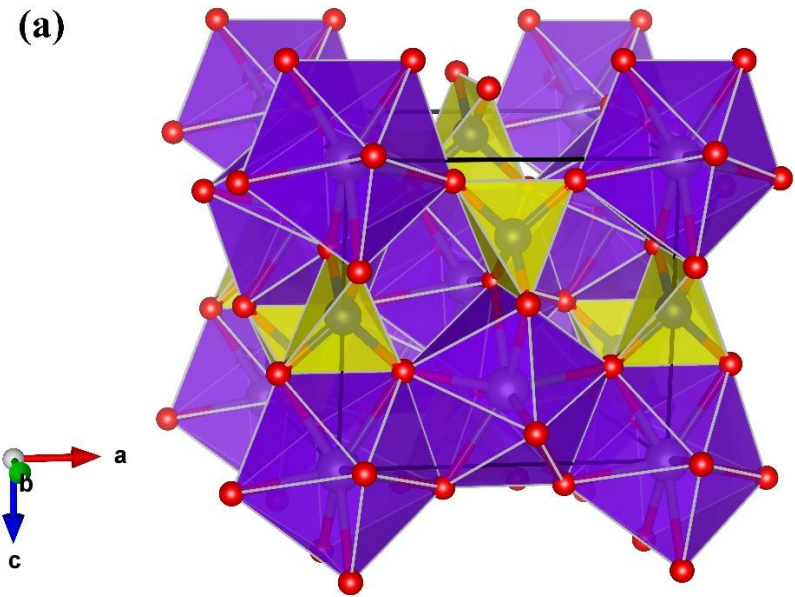

(a)

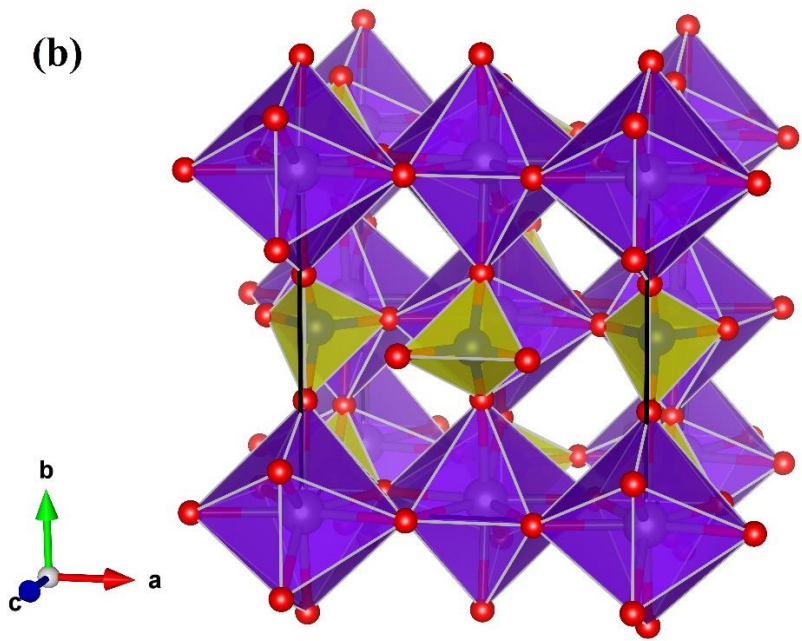

Figure 2. Lattice structure of HfSiO$_4$ unit cell. The figures (a) and (b) show views of (a,c) plane and (a,b) plane perspectives respectively. Oxygen atoms are shown in red; hafnium atoms are at the center of purple decahedra; silicon atoms are at the center of yellow tetrahedra. The decahedra and tetrahedra are edge-sharing. Each hafnium atom is connected to eight oxygen atoms, while every silicon atom has four oxygen neighbors. The primitive cell of HfSiO$_4$ has 12 atoms, while the conventional tetragonal cell has 24.

Table 1. Lattice parameters from simulations and experiments. Relative difference between prediction and the experimental values are shown in parentheses. The second experimental values followed by a (*) sign are from our own XRD measurements.

|  | PBEsol (present work) | Experiment [29] | LDA [24] (ABINIT) | LDA [27] (CASTEP) | GGA [28] (WIEN2k) | GGA [26] (CASTEP) | Empirical potential [25] |
|---|---|---|---|---|---|---|---|
| $a$ (Å) | 6.5822 (0.15%) | 6.5725 6.571(*) | 6.61 (0.57%) | 6.68 (1.64%) | 6.64 (1.03%) | 6.76 (2.85%) | 6.48 (1.41%) |
| $c$ (Å) | 5.9665 (0.06%) | 5.9632 5.972(*) | 5.97 (0.11%) | 5.97 (0.11%) | 6.08 (1.96%) | 6.04 (1.29%) | 6.06 (1.62%) |
| *u* | 0.0651 (0.61%) | 0.0655 | 0.0672 (2.60%) | 0.06939 (5.94%) | 0.07 (6.87%) | 0.0691 (5.50%) | 0.0070 (6.87%) |
| *v* | 0.1931 (0.87%) | 0.1948 | 0.1964 (0.82%) | 0.1977 (1.49%) | 0.19 (2.46%) | 0.1961 (0.67%) | 0.207 (6.26%) |

## 3.2 Mechanical properties

The tetragonal symmetry of hafnon dictates that there are six independent second-order elastic stiffness constants. Their predicted values, determined by the present study, together with other moduli derived from them, are compared with those obtained by various other methods in Table 2. The substantial differences in the elastic constants obtained by the different first-principles calculations is a result of the use of different exchange-correlation functionals. With the exception of $c_{13}$ and $c_{33}$, the PBEsol method used here gives elastic constants that are quite close to those predicted using LDA, while the GGA method predicts larger values. Under the Voigt-Reuss-Hill approximation, the Voigt bulk modulus and the Voigt shear modulus predicted by GGA have larger magnitude than those calculated from PBEsol, which is similar to the elastic constant coefficients. However, the Hill shear modulus predicted by the PBEsol, LDA and GGA methods are very similar. The Hill Young's modulus and Poisson's ratio obtained by PBEsol, LDA and GGA are all within 5% of the experimental values. The reason for this good agreement for Young's modulus obtained by the three methods is that for hafnon, the calculated Hill shear modulus is much smaller than the Hill bulk modulus, and this mainly determines the magnitude of the Hill Young's Modulus. The nanoindentation measurements yielded an averaged Young's elastic modulus of 297 ± 3% GPa. Its measured hardness (not calculated here) was 19.2 ± 0.8 GPa.

Table 2. Voigt-Reuss-Hill bulk modulus, shear modulus, and Young's modulus in GPa, and Poisson's ratio. Relative errors of Hill Young's modulus are in parentheses compared to the experimental value from nanoindentation.

|  | PBEsol (present work) | LDA [27] (CASTEP) | GGA [26] (CASTEP) | Empirical potential [25] | GGA [28] (WIEN2k) |
|---|---|---|---|---|---|
| $c_{11} = c_{22}$ | 430 | 484 | 604.8 | 441 | - |
| $c_{12}$ | 70 | 66 | 125.3 | 77 | - |
| $c_{13} = c_{23}$ | 151 | 160 | 217.7 | 192 | - |
| $c_{33}$ | 487 | 520 | 575.8 | 537 | - |
| $c_{44} = c_{55}$ | 110 | 106 | 103.2 | 107 | - |
| $c_{66}$ | 50 | 43 | 32 | 41 | - |
| Voigt bulk modulus | 233 | - | 323 | 260 | 236.72 |
| Voigt shear modulus | 119 | - | 129.3 | - | - |
| Voigt Young modulus | 306 | - | 342.2 | - | - |
| Voigt Poisson ratio | 0.28 | - | 0.323 | - | - |
| Reuss bulk modulus | 228 | - | 322.1 | - | - |
| Reuss shear modulus | 98.6 | - | 82.3 | - | - |
| Reuss Young modulus | 258 | - | 227.5 | - | - |
| Reuss Poisson ratio | 0.31 | - | 0.382 | - | - |
| Hill bulk Modulus | 230 | 249 | 322.5 | - | - |
| Hill shear modulus | 109 | 109 | 105.8 | - | - |
| Hill Young's modulus | 282 (5.0%) | 285 (4.0%) | 286.1 (3.7%) | - | - |
| Hill Poisson's ratio | 0.30 | 0.31 | 0.352 | - | - |
| The experimental Young's modulus deduced by nanoindentation (present work) was 297 ± 8.5 GPa while the hardness (present work) was 19.18 ± 0.78 GPa ||||||

### 3.3 Phonon dispersion and Gruneisen parameters

The phonon dispersion curves of hafnon within the Brillouin zone (BZ) of the primitive cell is displayed in Figure 3(a). Phonon frequencies determined by PBEsol (red curves) agree well with experimental data obtained by Raman spectroscopic techniques [19–21] and with results obtained from other first-principles calculations [24]. The MD results obtained using an empirical potential [25] differ slightly from our calculation along the $\Gamma - Z$ direction. The Born charge correction yields the LO-TO mode splitting and results in the frequency of 32.8 THz at the Γ point that is also predicted by the first-principles and the empirical MD calculations.

The phonon dispersion describes the harmonic properties of the hafnon crystal. In order to evaluate its anharmonicity, Gruneisen parameters that define the change in phonon frequency with change in unit cell volume were calculated. Figure 3(b) shows our calculated mode-Gruneisen parameters for hafnon using the ALAMODE and ShengBTE packages. The three groups of Gruneisen parameters match well, especially in the high frequency region. The difference near 5 THz may come from supercell size and cubic FCs fitting. For simplicity, we used 1×1×1 conventional cells for ShengBTE, and 2×2×2 supercells for ALAMODE calculations.

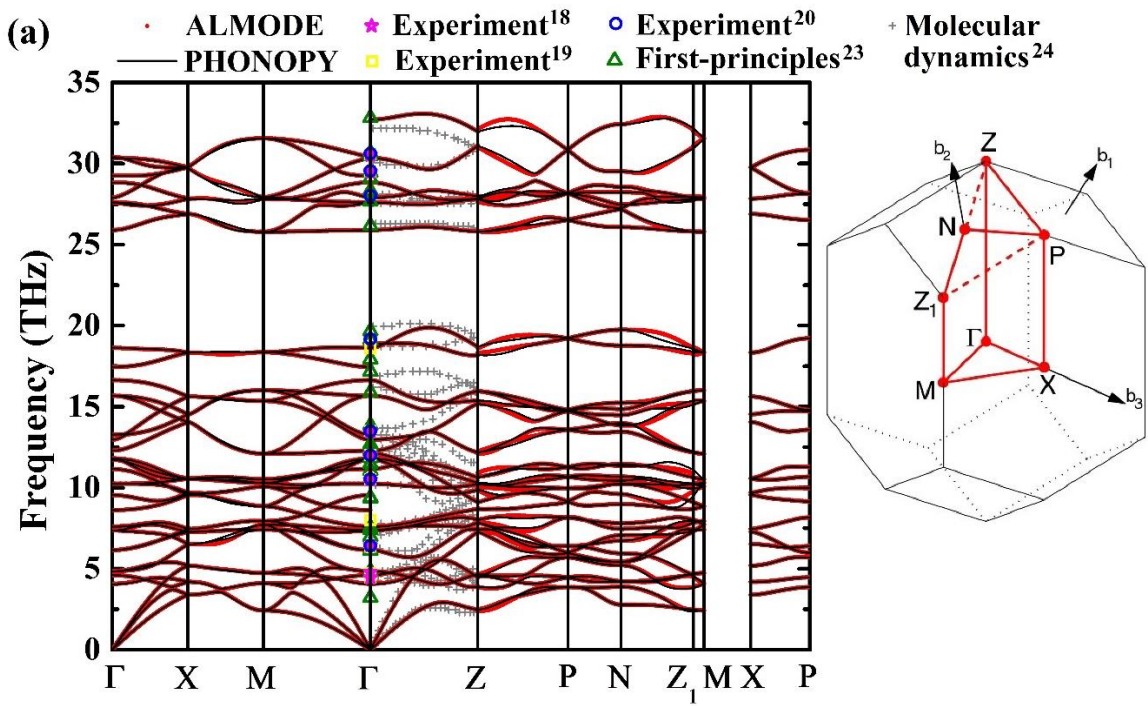

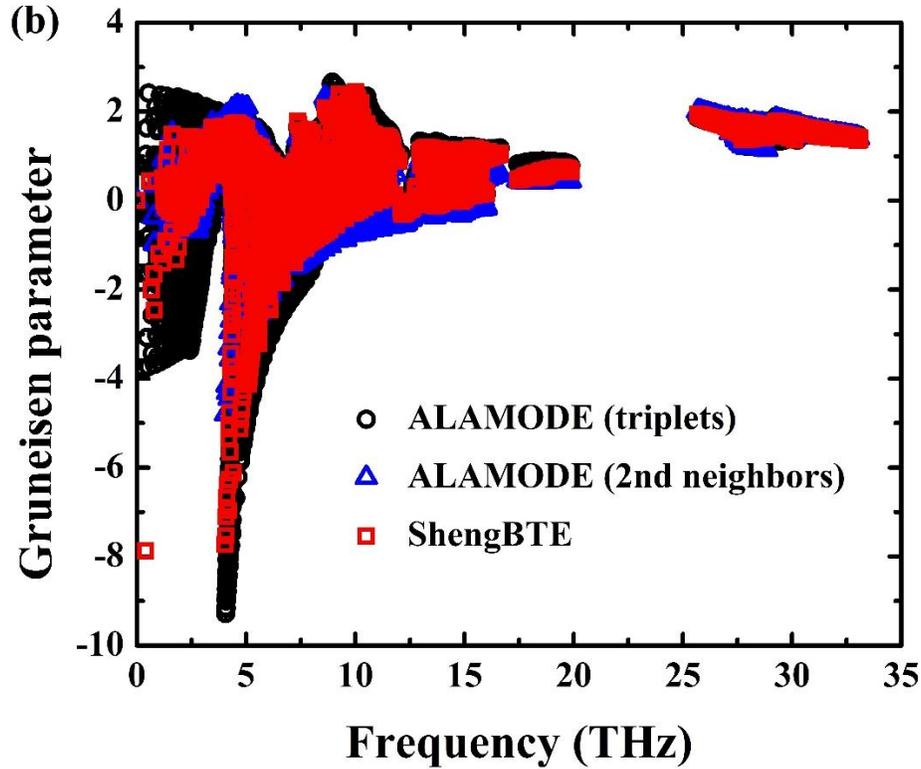

Figure 3. (a) Phonon dispersion versus wavevector of hafnon along high-symmetry directions in the first Brillouin zone of the primitive cell [49] shown on the right. $b_1$, $b_2$ and $b_3$ are reciprocal lattice vectors of the primitive cell. (b) Mode Gruneisen parameters calculated by the cubic FCs approximation using ALAMODE and ShengBTE. Nearest neighbors considered: the second nearest neighbors ($2^{nd}$ neighbors and ShengBTE) and all the triplets connected by bonds (triplets). Harmonic FCs are the same in the three groups.

### 3.4 Coefficients of thermal expansion

The measured and calculated temperature dependence of linear CTE is summarized in Figure 4. The XRD and the pushrod dilatometry measurements provide the linear CTE along the a- and c-axes, which we compare with our first-principles results. We also include results predicted by the

FP-LAPW method[28]. The simulation and the experimental results share the same temperature dependence and have similar orders of magnitude. Our first-principles calculations presented in this work yield closer values to the experiments. The linear CTE of hafnon is calculated using the average Gruneisen parameters, the compliance matrices and the volumetric CTE. The linear CTE along the c-axis is larger, which is consistent with the experiments, despite the difference in the high temperature region. In order to reduce this difference, we consider temperature effects on the compliance matrices. We keep the equilibrium volumes under high temperatures predicted by the QHA and adopt the ElaStic Code again to calculate the corrected compliances needed in the formula (3) that are listed in Table 3. The black squares and the round dots in Figure 4 are the corrected linear CTE in which the temperature dependence of the compliance tensor has been taken into account within the QHA. The temperature correction to the CTE is minor, the results are however closer to the experiments.

Table 3. The compliances in $10^{-4}$/GPa under 0K, 600K and 1200K, calculated within the QHA.

|  | $S_{11}$ | $S_{12}$ | $S_{13}$ | $S_{33}$ |
| --- | --- | --- | --- | --- |
| 0K | 26.05 | -1.6 | -7.57 | 25.2 |
| 600K | 27.65 | -1.51 | -8.18 | 26.43 |
| 1200K | 29.19 | -1.39 | -8.78 | 27.6 |

Table 4 shows our calculated, XRD and dilatometer measured CTE's and compares them with results from other groups. We believe that the dilatometer-measured negative CTE of $-1.8\times10^{-6}$ °C$^{-}$

[1] from 30 and 800 °C) in Ref. [50] could be due to the porosity of their samples. Other experiments [22,23] and simulations [27] have found the linear CTE of hafnon to lie in the range of $3.6 - 4.4\times10^{-6}$ $K^{-1}$ for temperatures between 25 and 1300 °C. These temperature-averaged values are consistent with our calculated and measured linear CTEs. For hafnon to be used as an EBC material, CTE should match that of Si bond coats and the SiC composites they protect. Since the linear CTE of Si is $3.5 – 4.5\times10^{-6}\,K^{-1}$ [51] and that of SiC/SiC melt infiltrated CMCs is $4.5 – 5.5\times10^{-6}\,K^{-1}$ [52], hafnon has a very good thermal expansion match with silicon EBC bond coats and the CMC substrates to which they are applied.

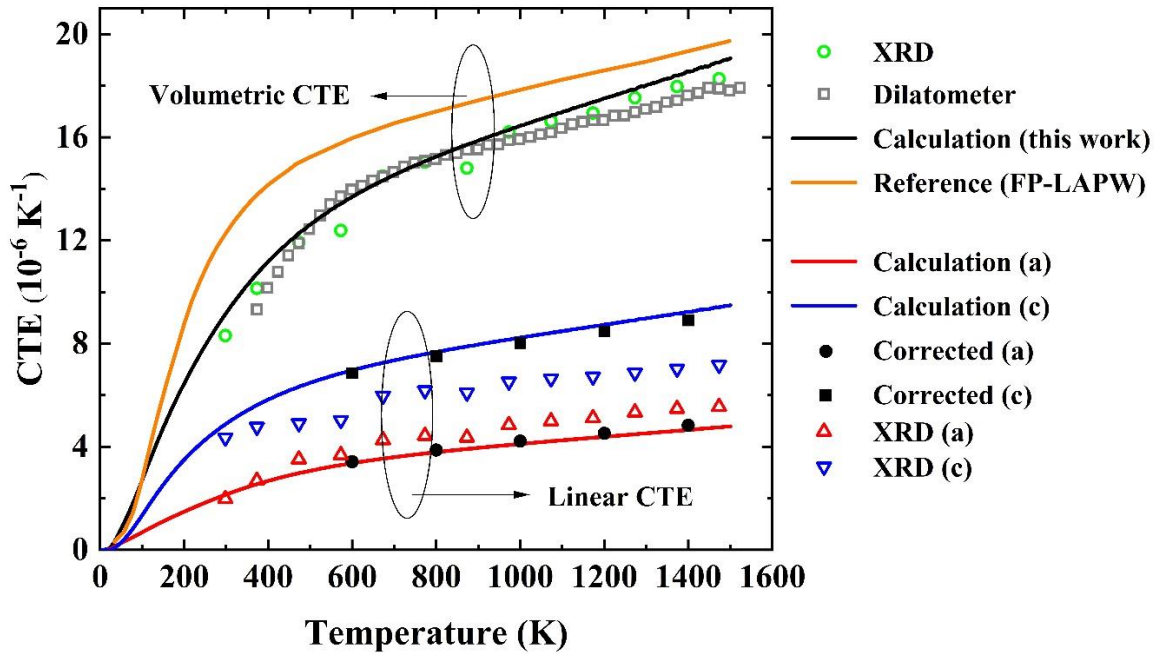

Figure 4. The temperature dependence of the linear and volumetric CTEs of hafnon. The corrected data (black circles and squares) include the temperature dependence of the compliance tensor.

Table 4. Averaged linear CTE from calculations and experiments (Unit: $10^{-6}$ K$^{-1}$). The chemical bond theory (CBT) result is displayed in the last column.

|  | Present calculation | XRD | Dilatometry | XRD [22,23] | XRD [50] | Dilatometry [50] | CBT [27] |
|---|---|---|---|---|---|---|---|
| Temperature (°C) | 27 – 1227 | 25 – 1200 | 100 – 1250 | 25 – 1300 | 25 – 800 | 30 – 800 | - |
| CTE | 3.06 – 6.36 | 2.77 – 6.09 | 3.11 – 5.97 | 3.6 | 4.17 | -1.8 | 4.41 |

## 3.6 Thermal transport properties

Figure 5(a) shows the thermal diffusivity and the thermal conductivity of hafnon calculated *via* the relaxation time approximation and measured by hot disk and laser flash methods. To calculate the thermal diffusivity, we used the specific heat and density from the QHA, as well as the thermal conductivity from the solution of the Boltzmann Transport Equation (BTE). To obtain converged thermal conductivity results, three meshes 16×16×16, 20×20×20 and 24×24×24 were used for sampling the first Brillouin zone. At each temperature, the reported thermal conductivity, κ was obtained by extrapolating the results from the three mesh values to infinite mesh. Since hafnon is a tetragonal crystal, the thermal conductivity along the a- and b-axes are the same, consistent with the cumulative thermal conductivity shown in Figure 5(b).

As seen in Figure 5(a), the hot disk technique yields slightly larger thermal diffusivity than the laser flash approach. The difference between the experiments is negligible within the error bars. The thermal diffusivity was converted to thermal conductivity, using the theoretical specific heat obtained from the QHA. In the high temperature region, κ was proportional to the inverse temperature, consistent with three-phonon scattering. Using laser flash and the hot disk method, the thermal conductivity of hafnon was 13.3 W/m.K at 30 °C and decreased to 6.25 W/m.K at 500 °C. Experimental results by laser flash above 600 K show a similar trend and agree well with the calculation. From room temperature to 500 K, the laser flash values and the hot disk results are in good agreement, both slightly deviating from the power law of $\kappa \propto T^{-1}$. The slight porosity of the hafnon sample leads to a lower density compared to the theory. We therefore compare the thermal diffusivities between theory and experiment, since it is the latter that is directly measured. In the lower temperature region, the effect of the defect-phonon scattering becomes more important. As we found experimental evidence for a 2.4% concentration of hafnia in the samples, we assume that such impurity scattering contributes to a lowering of the thermal conductivity. To take this into account, we adopt the impurity scattering approach of Tamura[53] to model the scattering of phonons of hafnon with residual hafnia "nanoparticles". In this model, we have to define a "scattering parameter" $g$ through $g = \Sigma_i f_i \left(1 - \frac{m_i}{\bar{m}}\right)^2$ where $\bar{m}$ is the average mass and $m_i$ is the mass of species $i$ of concentration $f_i$. We assume two species, hafnon of concentration $f_1$ =97.6% and hafnia of concentration $f_2$ =2.4%, and take the deviation in the density instead of $\left(1 - \frac{m_i}{\bar{m}}\right)$ as the perturbation parameter. This scattering process is then added, without any fitting parameter, to the three-phonon scattering rates in order to calculate the total thermal conductivity. The reduction of $\kappa$ due to this additional scattering is displayed in Fig. 5(a). Assuming Hafnia to have the bulk thermal conductivity of 1.1 W/mK, a mean thermal conductivity weighted by the concentration would only lead to only a 1% decrease in the thermal conductivity, while the impurity scattering model leads to a lowering of almost 10% at room temperature. We believe the scattering approach adopted above is a more realistic approach, as it also includes the proper temperature dependence of the scattering rates.

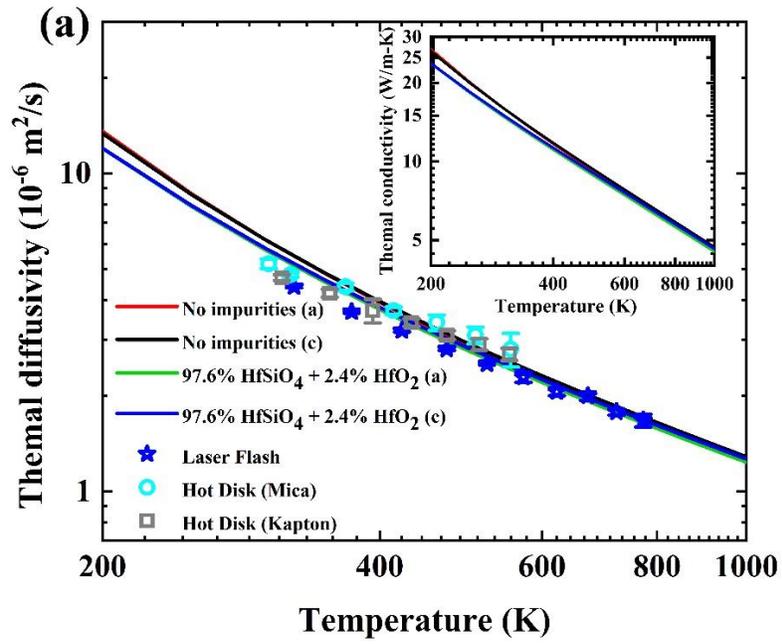

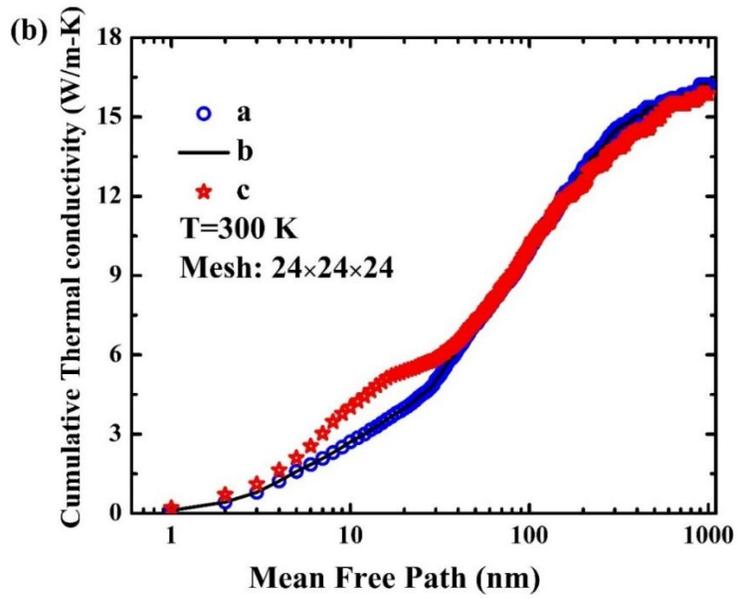

Figure 5. (a) The thermal diffusivity of hafnon from anharmonic calculations, laser flash and hot disk measurements. Kapton and Mica sensors distinguish the two hot disk measurements. The inset shows the calculated thermal conductivity of hafnon versus temperature in a Log-Log plot to show the inverse temperature relation. (b) Cumulative thermal conductivities versus mean free path of phonons along the a-, b- and c-axes at 300 K with the mesh size of 24×24×24.

## 4. Conclusions

Due to its possible applications as radioactive materials storage, EBC in the aerospace industry, and high-K dielectric in microelectronic industries, we undertook a comprehensive study of the physical, mechanical, and thermal transport properties of hafnon using both first-principles calculations and experiments. The first-principles methods were able to reliably predict crystal structure/lattice parameters, elastic constants, linear CTE and the thermal conductivity of hafnon. Amongst the first principles methods, PBEsol most accurately predicted the lattice parameters (to within 1%). The hafnon elastic constants were also well predicted by this method. The Hill Young modulus was within 5% of the experimental value obtained using a nanoindentation technique. The predicted linear CTE agreed well with both hot stage XRD lattice parameter and pushrod dilatometer measurements. The CTE of hafnon was found to be close to that of silicon. This makes it an attractive material for EBC applications. The predicted temperature-dependent thermal diffusivity and thermal conductivity were also in good agreement with experimental values obtained using hot disk and laser flash methods. It was shown that a concentration of hafnia as small as 2.5% can contribute to a lowering of room-temperature thermal conductivity by slightly less than 10%. Such strategies can me used to further lower the thermal conductivity of this material. The results of this study can be used to support further exploration of the above-mentioned applications.

## Acknowledgments

We are grateful for the research support of the University of Virginia's Research Innovation Awards initiative from the School of Engineering and Applied Sciences, and in JD and HW's case by ONR under Grant Numbers N00014-18-1-2645 and N00014-15-1-2756 managed by Dr. David Shifler.

## References

[1]     R.C. Ewing, W. Lutze, Zircon: A host-phase for the disposal of weapons plutonium, J. Mater. Res. 10 (1995) 243–246. https://doi.org/10.1557/JMR.1995.0243.

[2]     A. Meldrum, S.J. Zinkle, L.A. Boatner, R.C. Ewing, A transient liquid-like phase in the displacement cascades of zircon, hafnon and thorite, Nature. 395 (1998) 56–58. https://doi.org/10.1038/25698.

[3]     S.R. Levine, E.J. Opila, M.C. Halbig, J.D. Kiser, M. Singh, J.A. Salem, Evaluation of ultra-high temperature ceramics for aeropropulsion use, J. Eur. Ceram. Soc. 22 (2002) 2757–2767. https://doi.org/10.1016/S0955-2219(02)00140-1.

[4]     E. Wuchina, E. Opila, M. Opeka, W. Fahrenholtz, I. Talmy, UHTCs: Ultra-High Temperature Ceramic materials for extreme environment applications, Electrochem. Soc. Interface. 16 (2007) 30–36.

[5]     B.H. Lee, L. Kang, W.J. Qi, R. Nieh, Y. Jeon, K. Onishi, J.C. Lee, Ultrathin hafnium oxide with low leakage and excellent reliability for alternative gate dielectric application, in: Tech. Dig. - Int. Electron Devices Meet., IEEE, 1999: pp. 133–136. https://doi.org/10.1109/iedm.1999.823863.

[6]     G.D. Wilk, R.M. Wallace, J.M. Anthony, High-κ gate dielectrics: Current status and materials properties considerations, J. Appl. Phys. 89 (2001) 5243–5275. https://doi.org/10.1063/1.1361065.

[7]     D. Zhu, R.A. Miller, Sintering and creep behavior of plasma-sprayed zirconia- and hafnia-based thermal barrier coatings, Surf. Coatings Technol. 108–109 (1998) 114–120. https://doi.org/10.1016/S0257-8972(98)00669-0.

[8]     D. Zhu, R.A. Miller, Thermal conductivity and elastic modulus evolution of thermal

[8] barrier coatings under high heat flux conditions, J. Therm. Spray Technol. 9 (2000) 175–180. https://doi.org/10.1361/105996300770349890.

[9] D. Zhu, N. Bansal, K. Lee, Thermal conductivity of ceramic thermal barrier and environmental barrier coating materials, NASA TM-211122, NASA Glenn. (2001). http://www.ewp.rpi.edu/hartford/~lys/EP/Supporting Materials/References/zhu2001-2.pdf.

[10] D. Zhu, R.A. Miller, Development of Advanced Low Conductivity Thermal Barrier Coatings, Int. J. Appl. Ceram. Technol. 1 (2005) 86–94. https://doi.org/10.1111/j.1744-7402.2004.tb00158.x.

[11] C.G. Levi, Emerging materials and processes for thermal barrier systems, Curr. Opin. Solid State Mater. Sci. 8 (2004) 77–91. https://doi.org/10.1016/j.cossms.2004.03.009.

[12] D. Zhu, R.A. Miller, Hafnia-Based Materials Developed for Advanced Thermal/Environmental Barrier Coating Applications, NASA Rep. TM2005-192261, NASA, Clevel. (2004). https://ntrs.nasa.gov/search.jsp?R=20050192261.

[13] S. Ueno, D.D. Jayaseelan, H. Kita, T. Ohji, H.T. Lin, Comparison of Water Vapor Corrosion Behaviors of $Ln_2Si_2O_7$ (Ln=Yb and Lu) and $ASiO_4$ (A=Ti, Zr and Hf) EBC's, in: Key Eng. Mater., 2006: pp. 557–560. https://doi.org/10.4028/www.scientific.net/kem.317-318.557.

[14] T. Hisamatsu, I. Yuri, S. Ueno, T. Ohji, S. Kanzaki, Environmental barrier coating material and coating structure and ceramic structure using the same, 2006.

[15] G.D. Wilk, R.M. Wallace, J.M. Anthony, Hafnium and zirconium silicates for advanced gate dielectrics, J. Appl. Phys. 87 (2000) 484–492. https://doi.org/10.1063/1.371888.

[16] G.D. Wilk, R.M. Wallace, J.M. Anthony, Hafnium and zirconium silicates for advanced gate dielectrics, J. Appl. Phys. 87 (2000) 484–492. https://doi.org/10.1063/1.371888.

[17] J. Robertson, Band offsets of wide-band-gap oxides and implications for future electronic devices, J. Vac. Sci. Technol. B Microelectron. Nanom. Struct. 18 (2000) 1785–1791. https://doi.org/10.1116/1.591472.

[18] A. Callegari, E. Cartier, M. Gribelyuk, H.F. Okorn-Schmidt, T. Zabel, Physical and electrical characterization of Hafnium oxide and Hafnium silicate sputtered films, J. Appl. Phys. 90 (2001) 6466–6475. https://doi.org/10.1063/1.1417991.

[19] J.H. Nicola, H.N. Rutt, A comparative study of zircon ($ZrSiO_4$) and hafnon ($HfSiO_4$) Raman spectra, J. Phys. C Solid State Phys. 7 (1974) 1381–1386. https://doi.org/10.1088/0022-3719/7/7/029.

[20] P.W.O. Hoskin, K.A. Rodgers, Raman spectral shift in the isomorphous series ($Zr_{1-x}Hf_x$)$SiO4$, Eur. J. Solid State Inorg. Chem. 33 (1996) 1111–1121.

[21] B. Manoun, R.T. Downs, S.K. Saxena, A high-pressure Raman spectroscopic study of hafnon, $HfSiO4$, Am. Mineral. 91 (2006) 1888–1892. https://doi.org/10.2138/am.2006.2070.

[22] C.E. CURTIS, L.M. DONEY, J.R. JOHNSON, Some Properties of Hafnium Oxide,

Hafnium Silicate, Calcium Hafnate, and Hafnium Carbide, J. Am. Ceram. Soc. 37 (1954) 458–465. https://doi.org/10.1111/j.1151-2916.1954.tb13977.x.

[23] K. Nakano, N. Fukatsu, Y. Kanno, Thermodynamics of Zr/Hf-mixed silicates as a potential for environmental barrier coatings for Tyranno-hex materials, Surf. Coatings Technol. 203 (2009) 1997–2002. https://doi.org/10.1016/j.surfcoat.2009.01.035.

[24] G.M. Rignanese, X. Gonze, G. Jun, K. Cho, A. Pasquarello, Erratum: First-principles investigation of high-κ dielectrics: Comparison between the silicates and oxides of hafnium and zirconium (Physical Review B (2004) 69 (184301)), Phys. Rev. B - Condens. Matter Mater. Phys. 70 (2004) 099903. https://doi.org/10.1103/PhysRevB.70.099903.

[25] P.P. Bose, R. Mittal, S.L. Chaplot, Lattice dynamics and high pressure phase stability of zircon structured natural silicates, Phys. Rev. B - Condens. Matter Mater. Phys. 79 (2009) 174301. https://doi.org/10.1103/PhysRevB.79.174301.

[26] Q.J. Liu, Z.T. Liu, L.P. Feng, H. Tian, W. Zeng, First-Principles Investigations on Structural, Elastic, Electronic, and Optical Properties of Tetragonal HfSiO 4, Brazilian J. Phys. 42 (2012) 20–27. https://doi.org/10.1007/s13538-012-0067-0.

[27] H. Xiang, Z. Feng, Z. Li, Y. Zhou, Theoretical investigations on mechanical and thermal properties of MSiO4 (M = Zr, Hf), J. Mater. Res. 30 (2015) 2030–2039. https://doi.org/10.1557/jmr.2015.172.

[28] F. Chiker, F. Boukabrine, H. Khachai, R. Khenata, C. Mathieu, S. Bin Omran, S. V. Syrotyuk, W.K. Ahmed, G. Murtaza, Investigating the Structural, Thermal, and Electronic Properties of the Zircon-Type ZrSiO 4 , ZrGeO 4 and HfSiO 4 Compounds, J. Electron. Mater. 45 (2016) 5811–5821. https://doi.org/10.1007/s11664-016-4767-z.

[29] J.A. Speer, B.J. Cooper, Crystal structure of synthetic hafnon, HfSiO4, comparison with zircon and the actinide orthosilicates., Am. Mineral. 67 (1982) 804–808.

[30] P. Giannozzi, S. Baroni, N. Bonini, M. Calandra, R. Car, C. Cavazzoni, D. Ceresoli, G.L. Chiarotti, M. Cococcioni, I. Dabo, A. Dal Corso, S. De Gironcoli, S. Fabris, G. Fratesi, R. Gebauer, U. Gerstmann, C. Gougoussis, A. Kokalj, M. Lazzeri, L. Martin-Samos, N. Marzari, F. Mauri, R. Mazzarello, S. Paolini, A. Pasquarello, L. Paulatto, C. Sbraccia, S. Scandolo, G. Sclauzero, A.P. Seitsonen, A. Smogunov, P. Umari, R.M. Wentzcovitch, QUANTUM ESPRESSO: A modular and open-source software project for quantum simulations of materials, J. Phys. Condens. Matter. 21 (2009) 395502. https://doi.org/10.1088/0953-8984/21/39/395502.

[31] P. Giannozzi, O. Andreussi, T. Brumme, O. Bunau, M. Buongiorno Nardelli, M. Calandra, R. Car, C. Cavazzoni, D. Ceresoli, M. Cococcioni, N. Colonna, I. Carnimeo, A. Dal Corso, S. De Gironcoli, P. Delugas, R.A. Distasio, A. Ferretti, A. Floris, G. Fratesi, G. Fugallo, R. Gebauer, U. Gerstmann, F. Giustino, T. Gorni, J. Jia, M. Kawamura, H.Y. Ko, A. Kokalj, E. Küçükbenli, M. Lazzeri, M. Marsili, N. Marzari, F. Mauri, N.L. Nguyen, H. V. Nguyen, A. Otero-De-La-Roza, L. Paulatto, S. Poncé, D. Rocca, R. Sabatini, B. Santra, M. Schlipf, A.P. Seitsonen, A. Smogunov, I. Timrov, T. Thonhauser, P. Umari, N. Vast, X. Wu, S. Baroni, Advanced capabilities for materials modelling with Quantum ESPRESSO, J. Phys. Condens. Matter. 29 (2017) 465901. https://doi.org/10.1088/1361-

648X/aa8f79.

[32] J.P. Perdew, A. Ruzsinszky, G.I. Csonka, O.A. Vydrov, G.E. Scuseria, L.A. Constantin, X. Zhou, K. Burke, Generalized gradient approximation for solids and their surfaces, Phys. Rev. Lett. 100 (2007) 136406. https://doi.org/10.1103/PhysRevLett.100.136406.

[33] K.F. Garrity, J.W. Bennett, K.M. Rabe, D. Vanderbilt, Pseudopotentials for high-throughput DFT calculations, Comput. Mater. Sci. 81 (2014) 446–452. https://doi.org/10.1016/j.commatsci.2013.08.053.

[34] J.D. Pack, H.J. Monkhorst, "special points for Brillouin-zone integrations"-a reply, Phys. Rev. B. 16 (1977) 1748–1749. https://doi.org/10.1103/PhysRevB.16.1748.

[35] A. Togo, I. Tanaka, First principles phonon calculations in materials science, Scr. Mater. 108 (2015) 1–5. https://doi.org/10.1016/j.scriptamat.2015.07.021.

[36] T. Tadano, Y. Gohda, S. Tsuneyuki, Anharmonic force constants extracted from first-principles molecular dynamics: Applications to heat transfer simulations, J. Phys. Condens. Matter. 26 (2014) 225402. https://doi.org/10.1088/0953-8984/26/22/225402.

[37] Y. Wang, J.J. Wang, W.Y. Wang, Z.G. Mei, S.L. Shang, L.Q. Chen, Z.K. Liu, A mixed-space approach to first-principles calculations of phonon frequencies for polar materials, J. Phys. Condens. Matter. 22 (2010) 202201. https://doi.org/10.1088/0953-8984/22/20/202201.

[38] W. Li, J. Carrete, N.A. Katcho, N. Mingo, ShengBTE: A solver of the Boltzmann transport equation for phonons, Comput. Phys. Commun. 185 (2014) 1747–1758. https://doi.org/10.1016/j.cpc.2014.02.015.

[39] A. Togo, L. Chaput, I. Tanaka, G. Hug, First-principles phonon calculations of thermal expansion in Ti 3SiC2, Ti3AlC2, and Ti 3GeC2, Phys. Rev. B - Condens. Matter Mater. Phys. 81 (2010) 1–6. https://doi.org/10.1103/PhysRevB.81.174301.

[40] R. Golesorkhtabar, P. Pavone, J. Spitaler, P. Puschnig, C. Draxl, ElaStic: A tool for calculating second-order elastic constants from first principles, Comput. Phys. Commun. 184 (2013) 1861–1873. https://doi.org/10.1016/j.cpc.2013.03.010.

[41] R. Hill, The elastic behaviour of a crystalline aggregate, Proc. Phys. Soc. Sect. A. 65 (1952) 349–354. https://doi.org/10.1088/0370-1298/65/5/307.

[42] Y. Gong, C.W. McDonough, A.L. Beitelshees, N. El Rouby, T.P. Hiltunen, J.R. O'Connell, S. Padmanabhan, T.Y. Langaee, K. Hall, S.O.F. Schmidt, R.W. Curry, J.G. Gums, K.M. Donner, K.K. Kontula, K.R. Bailey, E. Boerwinkle, A. Takahashi, T. Tanaka, M. Kubo, A.B. Chapman, S.T. Turner, C.J. Pepine, R.M. Cooper-DeHoff, J.A. Johnson, PTPRD gene associated with blood pressure response to atenolol and resistant hypertension, in: J. Hypertens., 4th ed., 2015: pp. 2278–2285. https://doi.org/10.1097/HJH.0000000000000714.

[43] M. Gustavsson, E. Karawacki, S.E. Gustafsson, Thermal conductivity, thermal diffusivity, and specific heat of thin samples from transient measurements with hot disk sensors, Rev. Sci. Instrum. 65 (1994) 3856–3859. https://doi.org/10.1063/1.1145178.

[44] S.E. Gustafsson, Transient plane source techniques for thermal conductivity and thermal diffusivity measurements of solid materials, Rev. Sci. Instrum. 62 (1991) 797–804. https://doi.org/10.1063/1.1142087.

[45] T.K. Chu, C.Y. Ho, Thermal Conductivity and Electrical Resistivity of Eight Selected AISI Stainless Steels, West Lafayette, Indiana, 1978. https://doi.org/10.1007/978-1-4615-9083-5_12.

[46] G.M. Pharr, An improved technique for determining hardness and elastic modulus using load and displacement sensing indentation experiments, J. Mater. Res. 7 (1992) 1564–1583. https://doi.org/10.1557/JMR.1992.1564.

[47] Mikio Fukuhara, Asao Sanpei, High Temperature-Elastic Moduli and Internal Dilational and Shear Frictions of Fused Quartz, Jpn. J. Appl. Phys. 33 (1994) 2890–2893. https://doi.org/10.1143/JJAP.33.2890/meta.

[48] B.D. Beake, J.F. Smith, High-temperature nanoindentation testing of fused silica and other materials, Philos. Mag. A Phys. Condens. Matter, Struct. Defects Mech. Prop. 82 (2002) 2179–2186. https://doi.org/10.1080/01418610208235727.

[49] W. Setyawan, S. Curtarolo, High-throughput electronic band structure calculations: Challenges and tools, Comput. Mater. Sci. 49 (2010) 299–312. https://doi.org/10.1016/j.commatsci.2010.05.010.

[50] J. Varghese, T. Joseph, K.P. Surendran, T.P.D. Rajan, M.T. Sebastian, Hafnium silicate: A new microwave dielectric ceramic with low thermal expansivity, Dalt. Trans. 44 (2015) 5146–5152. https://doi.org/10.1039/c4dt03367a.

[51] R. Hull, Thermal Properties, in: Dev. Agric. Eng., INSPEC, The Institution of Electrical Engineers, London, United Kingdom, 1986: pp. 34–40. https://doi.org/10.1016/B978-0-444-99523-0.50008-8.

[52] R.W. Olesinski, G.J. Abbaschian, The C-Si (Carbon-Silicon) System, Bull. Alloy Phase Diagrams. 5 (1984) 486–489.

[53] S.-I. Tamura, Isotope scattering, Phys. Rev. B. 27 (1983) 858–866. https://journals.aps.org/prb/pdf/10.1103/PhysRevB.27.858.